\documentclass[showpacs,amsmath,amssymb,aps,prd,twocolumn]{revtex4-1}

\usepackage{graphicx}% Include figure files
\usepackage{dcolumn}% Align table columns on decimal point
\usepackage{bm}% bold math

\begin{document}

%\preprint{APS/123-QED}
\preprint{IFUP-TH/2015-1}

\title{Instantons and Monopoles}

\author{Adriano Di Giacomo}
\affiliation{University of Pisa, Department of Physics and INFN, Sezione di Pisa, Largo B. Pontecorvo 3, 56127 PISA, ITALY}%Lines break automatically or can be forced with \\ 
\email{adriano.digiacomo@df.unipi.it}

\author{Masayasu Hasegawa}
\affiliation{Joint Institute for Nuclear Research, Bogoliubov Laboratory of Theoretical Physics, Dubna, Moscow Region, 141980, RUSSIAN Federation}
\altaffiliation{University of Parma, Department of Physics and INFN, Gruppo Collegato di Parma, Via G. P. Usberti 7/A, 43124 PARMA, ITALY}%Lines break automatically or can be forced with \\
\email{hasegawa@theor.jinr.ru}

\date{\today}% It is always \today, today,
             %  but any date may be explicitly specified

\begin{abstract}
This study is part of a research program aimed to investigate the relations between instantons, monopoles, and chiral symmetry breaking. Monopoles are important 3-dimensional topological configurations existing in QCD, which are believed to produce colour confinement. Instantons are 4-dimensional topological configurations and are known to be related to chiral symmetry breaking. To study the relation between monopoles and instantons we generate configurations adding to the vacuum state static monopole-antimonopole pairs of opposite charges by use of a monopole creation operator. We observe that the monopole creation operator only adds long monopole loops to the configurations. We then count the number of fermion zero modes using Overlap fermions as a tool. As a result we find that each monopole-antimonopole pair of magnetic charge one adds one zero mode of chirality $\pm1$, i.e. one instanton of topological charge $\pm1$.
\end{abstract}

%\begin{description}
%\item[PACS numbers] 11.15.Ha, 12.38.Gc
%\end{description}
%\end{abstract}

% PACS, the Physics and Astronomy
                             % Classification Scheme.
%\keywords{Suggested keywords}%Use showkeys class option if keyword
                              %display desired
\maketitle
%\tableofcontents

\section{Introduction}

 Monopoles are 3-dimensional topological QCD configurations, which are most probably responsible for colour confinement, by condensing in the vacuum and producing dual superconductivity~\cite{'tH2}. This mechanism has extensively been studied on the Lattice during the years with different approaches. 
  
 Instantons are 4-dimensional configurations and are related to Chiral symmetry breaking as described e.g. by the instanton liquid model~\cite{Shuryak1}. 
  
 There exist strong hints that the role of the two configurations can be related. In Ref.~\cite{Rub} it is shown that monopoles catalyse instantons, in the sense that the probability of having an instanton configuration in the presence of a monopole is of order $1$. The analysis there was specifically addressed to grand-unified models in view of the proton decay, but it is equally valid in $QCD$.  On the other hand the work of Ref.~\cite{KvB, Dyak} shows that instantons, and more generally calorons are made of monopoles.
  
 As a first step towards a quantitative understanding of this relation, we add static monopole-antimonopole pairs to the vacuum of an $SU(3)$ pure gauge theory~\cite{DiGiacomo1, DiGiacomo2, DiGiacomo3}, with the aim of changing the number of monopoles without significant modifications of the vacuum state. Indeed, if the vacuum is a dual superconductor adding a pair of magnetic charges does not modify the ground state, which is an eigenstate of the creation operators, except in the total number of positive and negative monopoles. Magnetic charges are in any case shielded and the additional monopoles are one-dimensional structures propagating in time: their effect on the vacuum vanishes at large volumes as $V^{-\frac{3}{4}}$. Moreover the total magnetic charge being zero the operator which creates the monopoles has no infrared problems~\cite{DiGiacomo3}. On the states constructed in this way we measure the change in the number of instantons.
  
 As a tool to count instantons we use the zero modes of an Overlap Dirac Operator. In doing that we realise that each lattice configuration contains zero modes of only one sign, either $+$ or $-$, and never zero modes of opposite sign at the same time: this contrasts with the naive expectation that, zero modes being localised and the correlation length finite, independent parts of the lattice could contain instantons of opposite signs. We also would expect that the number of instantons and of anti-instantons, i.e. the number of zero modes, be proportional to the volume. We argue that, for some reason to be quantitatively studied and understood we only observe the net topological charge, while pairs of zero modes of opposite sign escape detection. Similar views can be found in~\cite{HKN}. Maybe this phenomenon is due to our rather small volumes, and will disappear at larger volumes. We check this argument by doing a careful study of the distributions of topological charges, a careful determination of the topological susceptibility, which we compare with existing results in the literature, to convince ourselves that the determination of the topological charge is correct and also that our code is correct [Section II].

 The number of instantons can in any case be determined from the topological susceptibility under the very general assumptions of translation invariance, existence of a finite correlation length, and  $CP$ invariance [Section III].
 
 In Section IV we briefly recall the definition of the operator which creates a static pair of monopoles propagating in time from $t= - \infty$ to $t= + \infty$ and sitting at a given spatial distance.

 In Section V we check that the operator really creates monopoles by using the technique of Ref.~\cite{Kanazawa1, Kanazawa2}. We find that indeed it creates so called long monopoles, which are not short-range fluctuations but wrap the lattice along the time direction.

 Finally in Section VI we look at the configurations with monopole pairs of different magnetic charges added, we measure the topological charge and from it we compute the number of instantons. The result is that a pair of monopoles of charges $\pm 1$ produces one instanton or anti-instanton. A pair with magnetic charge $m_{c}$ produces $m_{c}$ instantons or anti-instantons.

 In Section VII we draw some conclusions. 

 Some of the results have already been reported at the LATTICE 2014 Conference~\cite{LAT} and at the Conference Confinement 2014~\cite{CON}.

\section{Overlap fermions}

 Wilson fermions explicitly break Chiral symmetry. Therefore we use Overlap fermions which preserve Chiral symmetry on the  Lattice. The Ginsparg-Wilson relation~\cite{Gins1} describes Chiral symmetry in Lattice gauge theory, 
\begin{equation}
\gamma_{5} D + D \gamma_{5} = aDR\gamma_{5}D. \label{gw}
\end{equation}
$a$ is the lattice spacing, D the Overlap Dirac operator, and R a parameter. The right hand side of this relation is non zero, because of the Nielsen-Ninomiya~\cite{NN} no go theorem. Multiplying both sides of Eq. (\ref{gw}) by the inverse of the Overlap Dirac operator, $D^{-1}$, gives
\begin{equation}
\gamma_{5} D^{-1} + D^{-1} \gamma_{5} = aR\gamma_{5},
\end{equation}
showing that Chiral symmetry breaking of the propagator $D^{-1}$ is an operator of $\mathcal{O}(a)$ vanishing in the continuum limit. The exact form of the Overlap Dirac operator is defined in Ref.~\cite{Neuberger1} in terms of the massless Wilson Dirac operator $D_{W}$. 
\begin{equation}
D = \frac{1}{Ra} \left( 1 + \frac{A}{\sqrt{A^{\dagger}A}}\right)
\end{equation}
$A = - M_{0} + a D_{W}$, $M_{0}$ is a parameter, $0 < M_{0} < 2$. $D_{W}$ is the massless Wilson Dirac operator defined as follows:  
\begin{equation}
D_{W} = \frac{1}{2}\left\{ \gamma_{\mu}(\nabla_{\mu}^{*} + \nabla_{\mu}) - a\nabla_{\mu}^{*}\nabla_{\mu}\right\}\label{eq:def_wilson1}
\end{equation}
\begin{equation}
[\nabla_{\mu}\psi](n) = \frac{1}{a}\left\{ U_{n,\mu}\psi(n+\hat{\mu}) - \psi(n) \right\}
\end{equation}
\begin{equation}
[\nabla_{\mu}^{*}\psi](n) = \frac{1}{a}\left\{ \psi(n) - U_{n-\hat{\mu}, \mu}^{\dagger}\psi(n-\hat{\mu}) \right\}
\end{equation}

\subsection{Lattice spacing}\label{sec:lattice_a1}

\begin{table*}
\caption{Determination of the lattice spacing. The lattice spacing $a^{(0)}$ is computed by the analytic function of~\cite{Necco1}. The lattice spacing $a^{(1)}$ is computed by our simulations of the $\bar q q$ potential ($n$ is the number of smearing steps and $\alpha$ is the weight factor of smearing, FR is the range of the fit) and $a^{(2)}$ from our determination of the string tension.}\label{a1}
\begin{center}
{\renewcommand{\arraystretch}{1.2}
\begin{tabular*}{\textwidth}{c @{\extracolsep{\fill}}ccccccccc} \hline\hline
$\beta$ & $a^{(0)}$ [fm] & $a^{(1)}$ [fm] & $a^{(2)}$ [fm] & $V$ & ($n$, $\alpha$) & FR ($R_{I}/a$) & $\chi^{2}/ndf$ & $N_{conf.}$\\ \hline
6.00 & $9.315\times10^{-2}$ & $9.39(9)\times10^{-2}$ & $9.93(8)\times10^{-2}$ & $18^{4}$ &(20, 0.5) & 0.9 - 7.0 & 4.3/4.0 & 440  \\ \hline\hline
\end{tabular*}
}
\end{center}
\end{table*}
 To fix the scale, we determine the lattice spacing by using of the analytic interpolation of Ref.~\cite{Necco1}. As a check we also measure the string tension and the $\bar{q} q$ potential from Wilson loops. We use APE smearing~\cite{APE1} of the spacial link variables to suppress excited states and fit the function $V(R) = \sigma \cdot R - \alpha/R + C$ to the potential $V(R)$. The distance $R$ between quark and anti-quark is improved to $R_{I}$ using the Green function as defined in Ref.~\cite{Necco1, Lucsher1}. The Jackknife method is used to estimate the statistical errors.

 The lattice spacing is then computed in two different ways (1) $a^{(1)}$: from $\alpha$ and Sommer scale $r_{0} = 0.5 \ [\mbox{fm}]$, and  (2) $a^{(2)}$: from the String tension $\sqrt{\sigma} = 440 \ [\mbox{MeV}]$  [Table \ref{a1}]. The results are reasonably consistent with the analytic interpolation. Therefore, we take the lattice spacing from the analytic interpolation~\cite{Necco1}, and use the Sommer scale $r_{0} = 0.5 \ [\mbox{fm}]$.

\subsection{Simulation details}

\begin{table*}
\caption{This table summarises the results of our simulations for the number of zero modes and the topological susceptibility. $\chi^{2}/ndf$ refers to the fit of the function $P_{Q} = \frac{e^{-\frac{Q^{2}}{2\langle Q^{2}\rangle}}}{\sqrt{2\pi\langle Q^{2}\rangle}}$ to the distributions of the topological charges.}\label{tb:Nzero_1}
{\renewcommand{\arraystretch}{1.2}
\begin{tabular*}{\textwidth}{c @{\extracolsep{\fill}}ccccccccc} \hline\hline
 $\beta$  &  $a/r_{0}$  & $V$ & $V/r_{0}^{4}$ & $N_{Zero}$ & $\langle Q^{2} \rangle $ & $\langle Q^{2} \rangle r_{0}^{4}/V $ & $\chi^{2}/ndf$ & $N_{conf.}$ \\ \hline
5.79   & 0.2795 & $12^{4}$          & 126.5 & 2.32(12) & 8.3(7)     & 0.066(6) & 4.7/14.0  & 200 \\   
5.81   & 0.2659 & $10^{4}$          & 49.96 & 1.44(4)  & 3.50(17)   & 0.070(3) & 8.5/11.0  & 844 \\ 
       &        & $14^{4}$          & 191.9 & 2.80(13) & 12.2(1.1)  & 0.064(5) & 12.4/17.0 & 249 \\ 
       &        & $16^{4}$          & 327.4 & 3.63(17) & 20.9(1.8)  & 0.064(6) & 19.6/19.0 & 275 \\ 
5.85   & 0.2484 & $12^{4}$          & 78.95 & 1.87(10) & 5.3(5)     & 0.068(6) & 3.9/11.0  & 200 \\ 
       &        & $16^{4}$          & 249.5 & 3.30(13) & 17.0(1.3)  & 0.068(5) & 22.0/19.0 & 344 \\ 
5.86   & 0.2395 & $14^{4}$          & 126.5 & 2.32(9)  & 8.3(7)     & 0.066(6) & 16.7/15.0 & 338 \\ 
5.90   & 0.2216 & $12^{4}$          & 49.96 & 1.40(4)  & 3.32(16)   & 0.067(3) & 10.1/11.0 & 835 \\ 
       &        & $16^{4}$          & 157.9 & 2.48(11) & 10.3(9)    & 0.065(5) & 18.1/19.0 & 320 \\ 
5.93   & 0.2129 & $14^{4}$          & 78.95 & 1.95(9)  & 5.8(5)     & 0.073(6) & 15.0/12.0 & 278 \\ 
5.99   & 0.1899 & $14^{4}$          & 49.96 & 1.39(4)  & 3.17(15)   & 0.063(3) & 8.8/11.0  & 862 \\ 
6.00   & 0.1863 & $12^{4}$          & 24.98 & 0.88(4)  & 1.47(11)   & 0.059(4) & 6.9/8.0   & 430 \\
       &        & $14^{4}$          & 46.28 & 1.35(5)  & 3.17(19)   & 0.069(4) & 10.2/11.0 & 592 \\
       &        & $12^{3}\times24$ & 49.96 & 1.34(4)  & 2.98(15)    & 0.060(3) & 11.7/12.0 & 790 \\  
       &        & $16^{4}$          & 78.95 & 1.81(7)  & 5.4(4)     & 0.068(5) & 7.0/14.0  & 405 \\
       &        & $18^{4}$          & 126.5 & 2.42(11) & 8.8(7)     & 0.069(5) & 16.7/14.0 & 258 \\ 
6.07   & 0.1662 & $16^{4}$          & 49.96 & 1.33(4)  & 3.04(15)   & 0.061(3) & 4.1/12.0  & 895 \\ \hline\hline
\end{tabular*}
}
\end{table*}

 We generate configurations using the Wilson gauge action and periodic boundary conditions. The number of iterations for the thermalization is $\mathcal{O}(2.0\times10^{4}$), and the configurations are sampled after $\mathcal{O}(5.0\times10^{3}$) iterations between them. The simulation parameters are listed in Table \ref{tb:Nzero_1}.

 We construct the Overlap Dirac operator $D$ from the massless Wilson fermions $D_{W}$ which are computed with the gauge links of the configurations. An anti-periodic boundary condition in the temporal direction is used for the Wilson fermions. 

 In more detail, we carry out the numerical computations by the technique of Ref.~\cite{Giusti1, Galletly1, Weinberg2}. The massless Overlap Dirac operator $D(\rho)$ is 
\begin{equation}
D(\rho) = \frac{\rho}{a} \left\{ 1 + \frac{D_{W}(\rho)}{\sqrt{D_{W}(\rho)^{\dagger}D_{W}(\rho)}}\right\}.
\end{equation}
$D_{W}(\rho)$ is computed from massless Wilson Dirac operator $D_{W}$, Eq. (\ref{eq:def_wilson1}), as follows: 
\begin{equation}
D_{W}(\rho) = D_{W} - \frac{\rho}{a}
\end{equation}
$\rho$ is a (negative) mass parameter, $0 < \rho < 2$. We set $\rho = 1.4$ in this study. Using the Hermitian Wilson Dirac operator $H_{W}(\rho) = \gamma_{5}D_{W}(\rho)$, the sign function is defined as in Ref.~\cite{Neuberger1}
\begin{equation}
\epsilon(H_{W}(\rho)) \equiv \frac{H_{W}(\rho)}{\sqrt{H_{W}(\rho)^{\dagger}H_{W}(\rho)}}.
\end{equation}
Therefore, the massless Overlap Dirac operator is computed by the sign function
\begin{equation}
D(\rho) = \frac{\rho}{a} \left\{ 1 +  \gamma_{5}\epsilon(H_{W}(\rho)) \right\}.
\end{equation}

 Next, we numerically compute the sign function by using the minmax approximation by Chebyshev polynomials of Ref.~\cite{Galletly1, Giusti1, Edwards1}. We solve eigenvalue problems by using the Arnoldi method (subroutines of ARPACK), and save $\mathcal{O}$(80) pairs of the low-lying eigenvalues and eigenvectors of the Overlap Dirac operator. 
 
 If $n_{+}$ is the number of exact zero modes of plus chirality and $n_{-}$ of minus chirality in the spectrum the massless Overlap Dirac operator, the topological charge is defined as $Q = n_{+} - n_{-}$. The average square of the topological charge $\langle Q^{2} \rangle$, and the topological susceptibility $\langle Q^{2} \rangle /V$ are computed from the topological charges. All of our results are listed in Table \ref{tb:Nzero_1}.

\subsection{Eigenvalues and Spectral density}

\begin{figure}[htbp]
 \includegraphics[width=85mm]{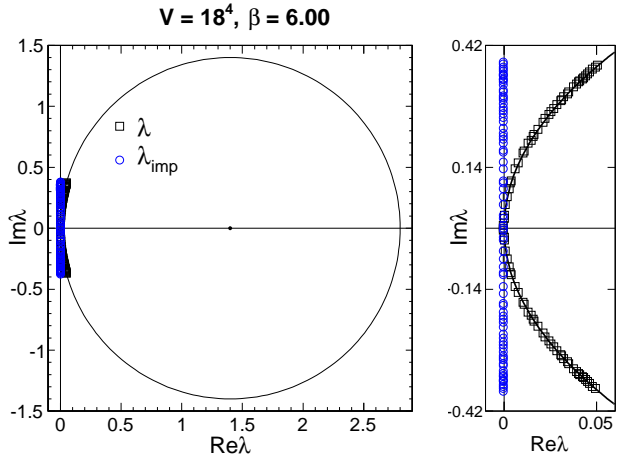}
 \caption{An example of distribution of eigenvalues $\lambda$, and improved eigenvalues $\lambda_{imp}$. In this configuration there is one zero mode. The right figure is an enlargement of the region $\lambda \approx 0$.}
 \label{fig:Dis_eigen1}
\end{figure}

\begin{figure}[htbp]
 \includegraphics[width=90mm]{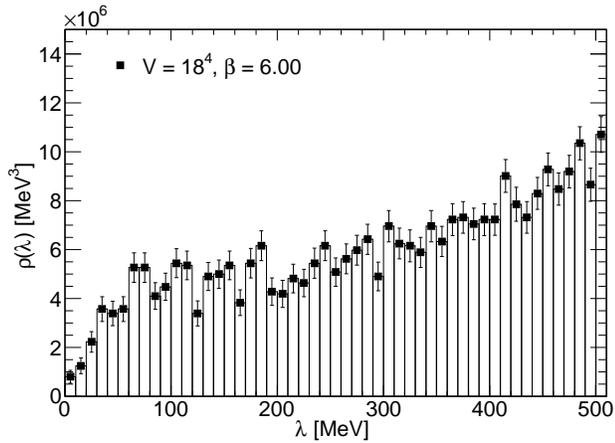}
 \caption{Spectral density of non zero modes ($\rho(\lambda)$).}
 \label{fig:Spec_dens2}
\end{figure}

 The eigenvalues $\lambda$ of the Overlap Dirac operator lie in the complex plane on a circle of centre $(\rho, 0)$ and radius $\rho$ as shown in Figure \ref{fig:Dis_eigen1}. In our case, $\rho = 1.4$. We consider instead the eigenvalues $\lambda_{imp}$ of the improved massless Overlap Dirac operator $D^{imp}(\rho)$ ~\cite{Capitani1}. They lie on the imaginary axis as shown in Figure \ref{fig:Dis_eigen1}. The improved massless Overlap Dirac operator $D^{imp}(\rho)$ is defined as:
\begin{equation}
D^{imp}(\rho) = \left( 1 - \frac{a}{2\rho} D(\rho) \right)^{-1} D(\rho)\label{imp_op1}
\end{equation}
The spectral density $\rho(\lambda, \ V)$ is defined as
\begin{equation}
\rho (\lambda, \ V) = \frac{1}{V}\langle \sum_{\lambda}\delta(\lambda - \bar{\lambda})\rangle.\label{eq:spec_1}
\end{equation}
 $\bar{\lambda} = \mbox{Im}({\lambda_{imp}})$. We show a typical spectral density $\rho (\lambda, \ V)$ of the non zero modes in Figure \ref{fig:Spec_dens2}.

\subsection{The number of zero modes, the topological charge, and the topological susceptibility}

\begin{figure}[htbp]
\includegraphics[width=90mm]{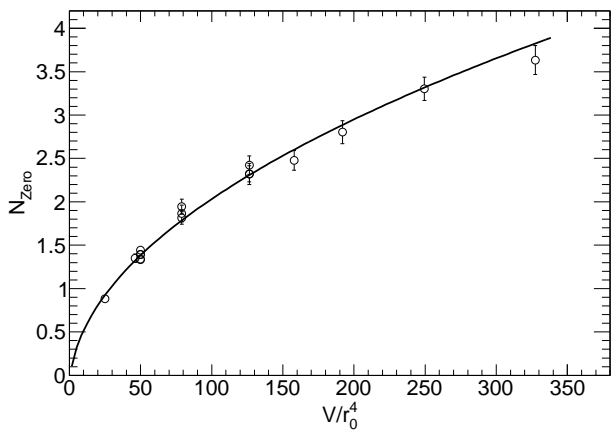}
\caption{The number of observed zero modes $N_{Zero}$ versus the physical volume $V/r_{0}^{4}$. The continuous curve is $N_{Zero} = \sqrt{A \cdot V/r_{0}^{4}} + B$, $A = 4.9(3)$, $B = -0.18(5)$.}\label{fig:Zero_allV1}
\end{figure}
\begin{figure}[htbp]
\includegraphics[width=90mm]{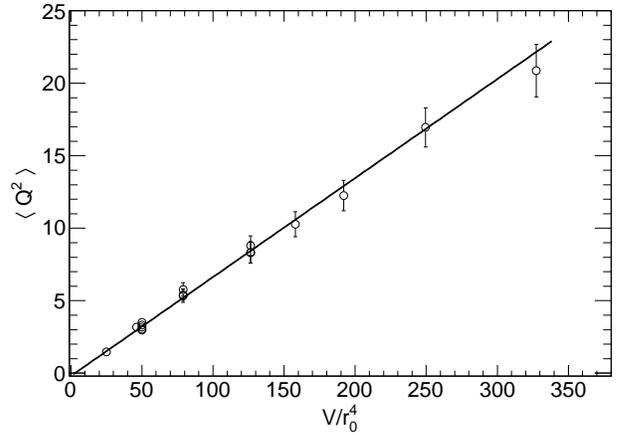}
\caption{The average square of the topological charges $\langle Q^{2}\rangle$ versus the physical volume $V/r_{0}^{4}$. The linear line is $\langle Q^{2}\rangle = A \cdot V/r_{0}^{4} + B$, $A = 6.8(2)\times10^{-2}$, $B = -0.20(13)$.}\label{fig:Q2_allv_2}
\end{figure}
 In our simulations, we never observe zero modes of opposite chirality in the same configuration. The zero modes have either all + chirality or all - chirality in each configuration: as a consequence the number $N_{obs}$ of observed zero modes  is $N_{obs} = | n_{+}-n_{-}| =|Q| $ and $N_{obs} ^2 = Q^2$. By definition $Q= n_{+} - n_{-}$.
 
 The average $N_{Zero} = \langle N_{obs} \rangle$ is plotted as a function of the volume $V$ in Figure \ref{fig:Zero_allV1} and is proportional to $\sqrt{V}$: fitting a function $N_{Zero} = \sqrt{A \cdot V/r_{0}^{4}} + B$ to the data gives $A = 4.9(3)\times10^{-2}$, $B = -0.18(5)$, and $\chi^{2}/ndf = 15.1/15.0$. The fitting range in physical volume units $V/r_{0}^{4}$ is from 24 to 330.  $\langle Q^2 \rangle$ is consequently linear in $V$ [Figure \ref{fig:Q2_allv_2}]. The slope is the topological susceptibility which is determined to be $\langle Q^{2} \rangle r_{0}^{4}/V = 6.8(2) \times 10^{-2}$. This value is consistent with the determination from the widths of the distributions of topological charges and with the determinations of other groups, as discussed below.
 
 However our counting of zero modes contrasts with the expectation that the number of zero modes be proportional to the volume. Indeed, if zero modes are associated to localized configurations like the instantons, and the vacuum is made of independent regions of size of the order of the correlation length, the number of instantons is proportional to the volume. In addition $\langle n_{+}\rangle = \langle n_{-}\rangle $ by invariance under $CP$. 
 
 If these arguments are correct the only way out is that pairs of instantons of opposite chirality somehow escape detection by our way of counting zero modes, at least at the rather modest volumes of our simulations. This effect preserves anyhow the value of the topological charge and would not affect the determination of the topological susceptibility. We shall use this assumption in the next Section to extract from our data the number density of instantons. A careful study of this phenomenon, however, should be done, to put on safer basis the determination of the topological susceptibility. A complementary strategy could be to improve numerically the determination based on the bosonic sector of the theory~\cite{add}, which does not rely on the counting of zero modes, and cross-check the two methods.

\begin{figure}[htbp]
  \includegraphics[width=90mm]{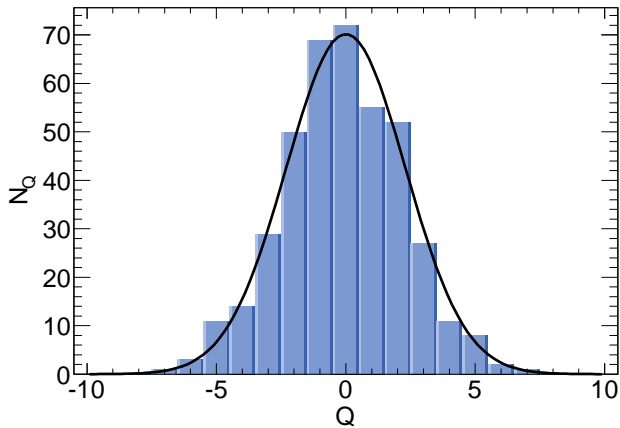}
\caption{A distribution of topological charges $Q$. The lattice is $V = 16^{4}$, $\beta = 6.00$.}
\label{fig:Dis_topQ1}
\end{figure}

 As anticipated above the topological susceptibility can be extracted from the distribution of topological charges. A typical distribution of the topological charges is that in Figure \ref{fig:Dis_topQ1}: a Gaussian function~\cite{Giusti2}
\begin{equation} 
P_{Q} = \frac{e^{-\frac{Q^{2}}{2\langle Q^{2}\rangle}}}{\sqrt{2\pi\langle Q^{2}\rangle}} \label{gauss_1}
\end{equation}
fits the data with $\chi^{2}/ndf = 7.0/14.0$, and topological susceptibility $\langle Q^{2}\rangle r_{0}^{4}/V = 6.7(5)\times10^{-2}$. This value is consistent with the value $\langle Q^{2}\rangle r_{0}^{4}/V = 6.8(5)\times10^{-2}$ directly computed from the number of zero modes. Moreover, seventeen distributions of the topological charges are computed from our seventeen different lattices, and we fit the Gauss function to all distributions. All the resulting values of $\chi^{2}/ndf$ are in the range from 0.3 to 1.3 [Table \ref{tb:Nzero_1}]. The topological charges all have the Gaussian distribution. 
\begin{figure}[htbp]
 \begin{center}
\includegraphics[width=90mm]{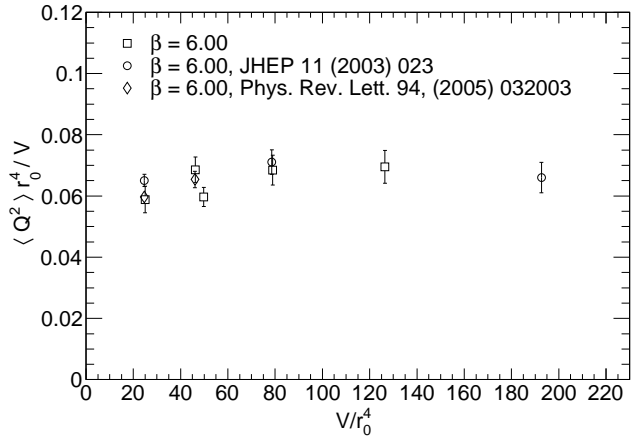}
\caption{The topological susceptibility $\langle Q^{2}\rangle r_{0}^{4}/V$ at $\beta = 6.00$ compared to results of other groups~\cite{Giusti2, Giusti3}.}
\label{fig:Tops_sus_b6p00}
 \end{center}
\end{figure}

\begin{figure}[htbp]
 \begin{center}
\includegraphics[width=90mm]{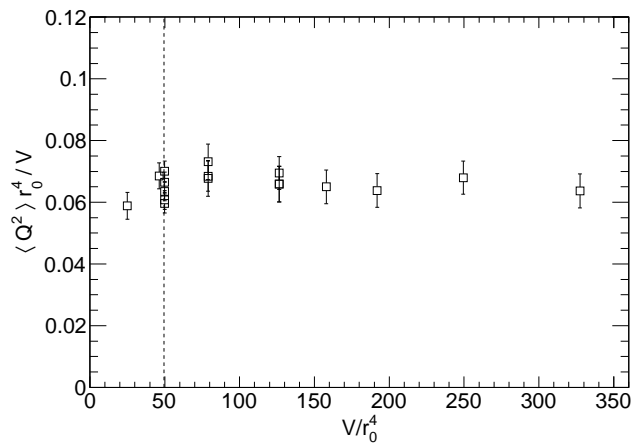}
\caption{The topological susceptibilities $\langle Q^{2}\rangle r_{0}^{4}/V$ in our simulations as shown in Table \ref{tb:Nzero_1} versus the physical volume $V/r_{0}^{4}$. A dotted line marks the physical volume $V/r^{4} = 50.00$.}
\label{fig:Top_sus_all1}
 \end{center}
\end{figure}

 In Figure \ref{fig:Tops_sus_b6p00}, we compare the topological susceptibility at the same $\beta = 6.00$ with results of other groups. The figure shows that our results are consistent with them. Moreover, we check that the finite lattice volume does not affect the topological susceptibility up to $L = 2.1 \ [\mbox{fm}]$ as indicated in Figure \ref{fig:Top_sus_all1}. 

\subsection{Topological susceptibility in the continuum limit}\label{sec:T_sus_con}

 Last, we fix the physical volume at the value $V/r_{0}^{4} = 50.00$ marked in Figure \ref{fig:Top_sus_all1}, and extrapolate the five data points of the topological susceptibility to the continuum limit using a linear expression $\langle Q^{2}\rangle r_{0}^{4}/V = c_{0} + c_{1} a^{2}$. The results are consistent with other groups. We are thus confident that eigenvalues and eigenvectors of overlap fermions in our simulations are properly computed.
\begin{flalign}
& \mbox{\textbf{Our result: }}  \chi  = (1.86 (6) \times 10^{2} \ [\mbox{MeV}])^{4}  \\
& \mbox{Ref~\cite{Giusti3}: } \chi  = (1.91 (5) \times 10^{2} \ [\mbox{MeV}])^{4} \\
& \mbox{Ref~\cite{DelDebbio1}: } \chi  = (1.88 (12) (5) \times 10^{2} \ [\mbox{MeV}])^{4} \\
& \mbox{The theoretical expectation~\cite{Veneziano1, Witten1}: }\nonumber\\
& \chi =\frac{\mbox {F}_{\pi}}{6} (m^{2}_{\eta} + m^{2}_{\eta'} - 2m_{K}^{2})|_{exp} \simeq (1.80\times 10^{2} \ [\mbox{MeV}])^{4}
\end{flalign}

\section{Instantons}\label{sec:inst}

 We want to compute the number of instantons from the number of zero modes, but, as anticipated, there are problems to determine it since some of them escape detection. In any translation invariant model indeed, e.g. the instanton liquid model, the number of instantons linearly increases with the physical volume. The number of our zero modes instead clearly increases as the square root of the physical volume [Figure \ref{fig:Zero_allV1}]. We never observe zero modes $n_{+}$ and $n_{-}$ of opposite chirality in the same configuration: our number of zero modes always coincides with the topological charge. The distributions of the topological charges determined in this way are Gaussian with $ 0.3 <  \chi^{2}/\mbox{ndf}  <  1.3$, and agree with other groups results.

 All that suggests that we observe the \textbf{net} number of zero modes. At least at our rather small physical lattice volumes for some reason pairs of zero modes of opposite sign seem to appear as non-zero modes. This can explain why we obtain the correct topological charge anyway. To estimate the density of instantons we can use a model based on the reasonable assumption that the instantons of both Chiralities are uniformly distributed in space-time and independent.

\subsection{The number of instantons}

 Let us denote the number of instantons of positive chirality in a volume $V$ by $n_{+}$, the number of instantons of negative chirality by $n_{-}$. Of course 
\begin{equation}
\langle n_{+} \rangle =\langle n_{-} \rangle = \frac{N_{I}}{2} = \rho_{i} V,
\end{equation}
$\rho_{i}$ is the density of instantons. Because of $CP$ invariance instantons and anti-instantons have the same distribution. The instantons are independent and the distribution Poisson-like $O(\frac{1}{V_{st}})$: indeed we can view a space-time of volume $V_{st}$ as covered by non overlapping hypercubes of size $V$, which are all equivalent by translation invariance, independent from each other if their size is bigger than the correlation length, and can all be used at the same time to determine the distribution, which is then Poisson-like.
\begin{equation}
\mbox{P}(n_{+}) = \frac{1}{n_{+}!}\left(\frac{N_{I}}{2}\right)^{n_{+}}\mathrm{e}^{\frac{-N_{I}}{2}}\label{n_{+}}
\end{equation}
\begin{equation}
\mbox{P}(n_{-}) = \frac{1}{n_{-}!}\left(\frac{N_{I}}{2}\right)^{n_{-}}\mathrm{e}^{\frac{-N_{I}}{2}} \label{n_{-}}
\end{equation}
The resulting distribution for $Q = n_{+} - n_{-} $ is
\begin{equation}
\mbox{P}(Q) = \sum^{\infty} _{n_{-}=0} P(n_{-}) P(n_{-}+Q) = \exp(-N_{I}) I_{Q}(N_{I}).
\end{equation}
Here $I_{Q} (x) $ is the modified Bessel function 
\begin{equation}
I_{Q} (x) = \sum_{k=0}^{\infty} (\frac{x}{2})^{n_{-}} (\frac{x}{2})^{n_{-}+Q}\exp(-N_{I}) \frac{1}{n_{-}! (n_{-} +Q)!}
\end{equation}
When $Q\gg 1$, $N_{I} \gg1$ at $\frac{Q^2}{N_{I}}$ fixed 
\begin{equation}
\mbox{P}(Q) \simeq \frac{\mathrm{e}^{-\frac{Q^{2}}{2N_{I}}}}{\sqrt{2N_{I}\pi}} \label{eq:pb_func1} 
\end{equation}
Finally, the number of instantons is determined as $N_{I} = \langle Q^{2} \rangle = \langle N_{Zero}^{2}\rangle$. 

\subsection{The density of instantons}\label{sec:ins1}

 We obtain the instanton density by fitting a linear function $ N_{I} = 2\rho_{i}r_{0}^{4} \cdot V/r_{0}^{4} + B$ to $\langle Q^{2} \rangle$ as shown in Figure \ref{fig:Q2_allv_2}. All of data points are included in the fit and $\chi^{2}/ndf = 11.9/15.0$. The slope is $2\rho_{i}r_{0}^{4} = 6.8(2)\times 10^{-2}$ and the intercept is $B = -0.20(13)$. The intercept is compatible with zero. Finally, the instanton density is evaluated as 
\begin{equation}
\rho_{i} = 8.3 (3) \times 10^{-4} \ [\mbox{GeV}^{4}].
\end{equation}
This result is consistent with the instanton liquid model, Ref.~\cite{Shuryak1}.

\section{The monopole creation operator}\label{sec:mon_cre1}

 In this section, we briefly recall the definition of the monopole creation operator and the method to count monopoles. In order to understand the relation between instantons and monopoles we add monopole-antimonopole pairs of opposite charges in the configurations using the monopole creation operator, and measure the variation of the number of instantons. We first check that the monopoles are successfully added in the configurations by counting the additional monopoles.

 The monopole creation operator is defined in~\cite{DiGiacomo1, DiGiacomo2, DiGiacomo3}. Specifically, in this study, we use the monopole creation operator defined in~\cite{DiGiacomo3}, [Eq. (41) et seq.], which, acting on the vacuum state, produces a static pair of monopole-anti-monopole of opposite charge propagating from $t = -\infty$ to $t = + \infty$. The operator is defined as
\begin{equation}
\bar \mu =\exp(-\beta \overline {\Delta S}) \label{m}.
\end{equation}
$\overline {\Delta S}$ is defined by modifying the normal action at the time $t$ by replacing the usual plaquette $\Pi_{\mu\nu}(n)$ by $\bar{\Pi}_{\mu\nu}(n)$, as follows:  
 \begin{equation} 
S + \overline{\Delta S} \equiv \sum_{n, \mu < \nu} \mbox{Re} (1 - \bar{\Pi}_{\mu\nu}(n))
\end{equation}
$\bar{\Pi}_{\mu\nu}(n) = \Pi_{\mu\nu}(n)$ in all sites $n$ with $n_0 \neq t$; at $n_0 = t$ the space-space components $i,j = 1 - 3$ are again unmodified $\bar{\Pi}_{i j}(t,\vec n) = \Pi_{i j}(t,\vec n)$ while $\bar{\Pi}_{i0}(t, \vec n)$ is a modified plaquette with inserted matrices $M_{i}(\vec{n})$ and $M_{i}(\vec{n})^{\dagger}$, 
\begin{align}
& \bar{\Pi}_{i0}(t, \vec{n}) = \frac{1}{\mbox{Tr}\mbox[I]}\mbox{Tr}[U_ {i}(t, \vec{n})M_{i}^{\dagger}(\vec{n} + \hat{i})\nonumber \\
& \times U_{0}(t, \vec{n} + \hat{i})M_ {i}(\vec{n} + \hat{i})U_{i}^{\dagger}(t + 1, \vec{n})U_{0}^{\dagger}(t, \vec{n})]
\end{align}
$\mbox{Tr}\mbox[I]$ is the trace of the identity, and the matrix $M_{i}(\vec{n})$ is the discretised version of the classical field configuration $\sum _{k} A_{i}^{0}(\vec{n} - \vec{x_{k}})$ produced at site $\vec {n}$ by the monopoles to be added in the locations $\vec x _{k}$, namely 
\begin{equation} 
M_{i}(\vec{n}) = \exp(i \sum _{k} A_{i}^{0}(\vec{n} - \vec{x_{k}})), \ ( i = x, \ y, \ z ).
\end{equation}
The form used for the monopole fields in a spherical coordinate system $(r, \ \theta, \ \phi )$ centred at the monopole is Wu-Yang:
\begin{align}
(i) \ n_{z} - z & \geqq 0  \nonumber \\
\begin{pmatrix}
A_{x}^{0} \\
A_{y}^{0} \\
A_{z}^{0}
\end{pmatrix}
& = 
\begin{pmatrix}
\frac{m_{c}}{2gr}\frac{\sin\phi (1 + \cos\theta)}{\sin\theta}\lambda_{3} \\
-\frac{m_{c}}{2gr}\frac{\cos\phi (1 + \cos\theta)}{\sin\theta}\lambda_{3} \\
0 
\end{pmatrix}
\end{align}

\begin{align}
(ii) \ n_{z} - z & < 0  \nonumber \\
\begin{pmatrix}
A_{x}^{0} \\
A_{y}^{0} \\
A_{z}^{0}
\end{pmatrix}
& = 
\begin{pmatrix}
-\frac{m_{c}}{2gr}\frac{\sin\phi (1 - \cos\theta)}{\sin\theta}\lambda_{3} \\
\frac{m_{c}}{2gr}\frac{\cos\phi (1 - \cos\theta)}{\sin\theta}\lambda_{3} \\
0 
\end{pmatrix}
\end{align}
The electric charge is
\begin{equation}
 g = \sqrt{\frac{6}{\beta}}: \ \mbox{(gauge coupling constant)}
\end{equation}
\textbf{We give the monopoles magnetic charges \boldmath${m_{c}= 0, 1, 2, 3, 4}$.} One monopole has charge $+m_{c}$ and the other has charge $-m_{c}$. The total magnetic charge is zero. $m_{c}=0$ is the reference configuration with no monopoles added.

 The monopole of charge $+m_{c}$ and anti-monopole of charge $-m_{c}$ are placed at time slice $t=T/2$: the choice is irrelevant since boundary conditions are periodic in time.

The locations of the monopoles in the lattice $V = L^{3}\times T$ are chosen in the $(x, y, z, t)$ space as in Table \ref{tb:location_mon_1}, where also the distance $N$ between the monopoles is defined. 
\begin{table}
\caption{The definition of Distance $N$ and the location of the monopole and the anti-monopole in the four-dimensional space $(x, y, z, t)$. $V = L^{3}\times T$.}\label{tb:location_mon_1}
{\renewcommand{\arraystretch}{1.3}
\begin{tabular*}{\columnwidth}{l @{\extracolsep{\fill}}ccc} \hline\hline
$N$ & Monopole & Anti-monopole \\ \hline
Odd  & $\left(\frac{L+N+2}{2}, \frac{L+N+2}{2}, \frac{L+1}{2}, \frac{T}{2}\right)$ & $\left(\frac{L-N+2}{2}, \frac{L-N+2}{2}, \frac{L-1}{2}, \frac{T}{2}\right)$ \\ 
Even & $\left(\frac{L+N+1}{2}, \frac{L+N+1}{2}, \frac{L+1}{2}, \frac{T}{2}\right)$ & $\left(\frac{L-N+1}{2}, \frac{L-N+1}{2}, \frac{L-1}{2}, \frac{T}{2}\right)$ \\ \hline\hline
\end{tabular*}
}
\end{table}

While Monte Carlo simulations are carried out, the pair of monopoles makes long monopole loops in the configurations. 
\section{Detecting the additional monopoles}\label{sec:ma_mono1}

 To verify that the monopoles are successfully added to the configurations, we detect the monopoles in the configurations, as done in Ref.~\cite{DeGrand1, Kanazawa1, Kanazawa2}. We generate the configurations varying the number of added monopole charges from zero to four. The distance between the monopole and the anti-monopole is defined in Table \ref{tb:location_mon_1}. Next, the configurations are iteratively transformed to the Maximally Abelian gauge using the Simulated Annealing algorithm. To remove the effects of the Gribov copies, 20 iterations are carried out in our simulations.

 Abelian link variables, holding $U(1)\times U(1)$ symmetry, are derived by Abelian projection from non-Abelian link variables. The monopole current is defined for each colour direction as follows:  
\begin{equation}
k_{\mu}^{i} (n) = \frac{1}{2}\epsilon_{\mu\nu\rho\sigma}\partial_{\nu}n_{\rho\sigma}^{i}(n+\nu)
\end{equation}
The colour index can be i = 1, 2, and 3. $n_{\rho\sigma}^{i}(n+\nu)$ is the Dirac string~\cite{Poly1}, and the monopole current satisfies the conservation law,
\begin{equation}
\sum_{i}k_{\mu}^{i} (n) = \sum_{i}\frac{1}{2}\epsilon_{\mu\nu\rho\sigma}\partial_{\nu}n_{\rho\sigma}^{i}(n+\nu) = 0.
\end{equation}
The monopole density is defined as follows:
\begin{equation}
\rho_{m}r_{0}^{3} = \frac{1}{12V}\sum_{i}\sum_{n, \mu}|k_{\mu}^{i}(n)| r_{0}^{3}.
\end{equation}
\begin{figure}[htbp]
  \begin{center}
    \includegraphics[width=90mm]{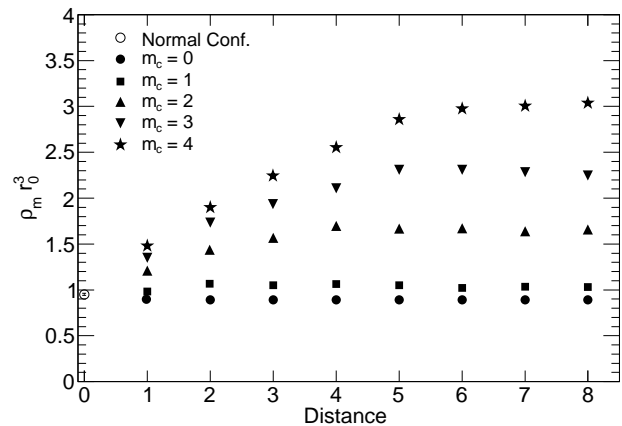}
    \caption{The monopole density $\rho_{m} r_{0}^{3}$ versus the distance between the monopole and anti-monopole. The lattice is $V=14^{4}, \beta = 6.00$.}\label{fig:distance_mo_antimon1}
 \end{center}
\end{figure}
 First, we measure the monopole density as shown in Figure \ref{fig:distance_mo_antimon1}, varying the distance between the monopole and anti-monopole from one to eight, moreover, varying the monopole charges $m_{c}$ from zero to four. The monopole densities go to plateaus when the monopole and anti-monopole are adequately separated. 
\begin{table}
\caption{The monopole density $\rho_{m} r_{0}^{3}$. D6 and D8 indicate Distance = 6 and 8 respectively. $N_{conf.}$ indicate the number of configurations each distance.}\label{tb:mon_dens_1}
{\renewcommand{\arraystretch}{1.2}
\begin{tabular*}{\columnwidth}{c @{\extracolsep{\fill}}ccccc} \hline\hline
$\beta$ & $V$ & $m_{c} $ & $\rho_{m} r_{0}^{3}$ (D6) & $\rho_{m} r_{0}^{3}$ (D8) & $N_{conf.}$ \\ \hline
6.00   & $14^{4}$ &    0    & 0.893(12)  & 0.893(12) & 80 \\
       &         &    1    & 1.026(12) & 1.032(13) & 80 \\
       &         &    2    & 1.658(14)  & 1.660(13) & 80 \\
       &         &    3    & 2.281(15)  & 2.247(17) & 80  \\
       &         &    4    & 2.991(14)  & 3.024(15) & 80  \\ 
6.00   & $18^{4}$ &    0   &  0.917(7)  & 0.916(9)  & 80 \\
       &         &    1    & 0.981(7) & 0.993(8)  & 80 \\
       &         &    2    & 1.307(7) & 1.305(8)  & 80 \\
       &         &    3    & 1.619(8) & 1.621(9)  & 80  \\
       &         &    4    & 1.998(9) & 2.056(8)  & 80  \\ \hline\hline
\end{tabular*}
}
\end{table}
 On this plateau we fix the distance D between the monopole and the anti-monopole, D = 6 and D=8, and measure the monopole density. The monopole density does not depend on the distance between them as indicated in Table \ref{tb:mon_dens_1}. The monopole densities computed from normal configurations of $\beta = 6.00$, $V = 14^{4}$ and $18^{4}$ are $\rho_{m} r_{0}^{3} = 0.940(11)$ and $\rho_{m} r_{0}^{3} = 0.919(7)$ respectively. 
\begin{figure}[htbp]
\includegraphics[width=90mm]{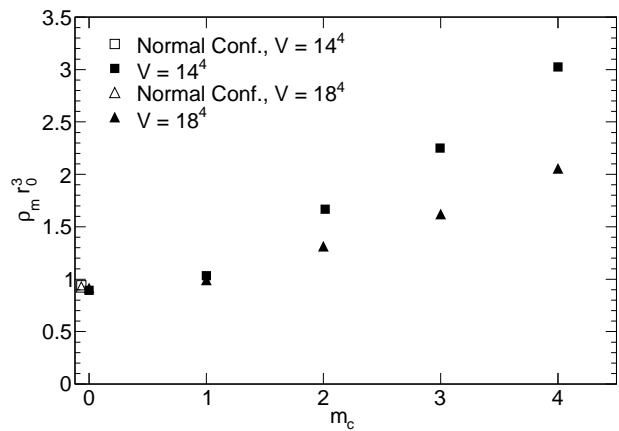}
\caption{The monopole density $\rho_{m} r_{0}^{3}$ versus the magnetic charge $m_c$. Distance = 8. The data points of the normal configurations are slightly shifted horizontally from $m_{c} = 0$ to distinguish them.}\label{fig:Mon_density_fig1}
\end{figure}
 The monopole density increases with monopole charges as shown in Figure \ref{fig:Mon_density_fig1}. However, if the physical volume becomes larger, the slope becomes lower.  As the physical volume becomes larger, the number density  of the monopoles is rarefied.  We add the number of monopoles in the configurations by increasing the monopole charges.

 Next, we measure the lattice spacing in the same way as in Section \ref{sec:lattice_a1}. The results are listed in Table \ref{add_a1} and show that the additional monopoles do not change it appreciably.

\begin{table*}
\caption{Determination of the lattice spacing. The lattice spacing $a^{(1)}$ and $a^{(2)}$ are computed by our simulations. $n$ is the number of smearing steps and $\alpha$ is the weight factor of smearing.}\label{add_a1}
{\renewcommand{\arraystretch}{1.2}
\begin{tabular*}{\textwidth}{c @{\extracolsep{\fill}}cccccccccc} \hline\hline
$\beta$ & $V$ & Distance & $m_{c} $ &  $a^{(1)}$ [fm] & $a^{(2)}$ [fm]  &($n$, $\alpha$) & FR ($R_{I}/a$) & $\chi^{2}/ndf$ & $N_{conf.}$\\ \hline
6.00  & $18^{4}$ & 6 & 0 & 0.092(3) & 0.0967(13) & (30, 0.5) & 1.8 - 8.0 & 4.0/4.0 & 500  \\
      &         &   & 1 & 0.0951(8) & 0.1005(7) & (25, 0.5) & 0.9 - 8.0 & 5.2/5.0 & 500  \\
      &         &   & 2 & 0.097(3) & 0.1021(17)  & (20, 0.5) & 1.8 - 8.0 & 3.8/4.0 & 500  \\
      &         &   & 3 & 0.097(4) & 0.101(2)  & (25, 0.5) & 1.8 - 7.0 & 2.9/3.0 & 500  \\
      &         &   & 4 & 0.1010(11) & 0.1068(10) & (25, 0.5) & 0.9 - 6.0 & 3.2/3.0 & 500  \\ 
      &         & 8 & 0 & 0.092(3) & 0.0968(14) & (30, 0.5) & 1.8 - 8.0 & 4.0/4.0 & 500  \\
      &         &   & 1 & 0.095(4) & 0.100(2) & (25, 0.5) & 1.8 - 6.0 & 2.0/2.0 & 500  \\
      &         &   & 2 & 0.0980(8) & 0.1036(7) & (25, 0.5) & 0.9 - 9.0 & 6.1/6.0 & 500  \\
      &         &   & 3 & 0.099(4) &  0.104(2) & (15, 0.5) & 1.8 - 7.0 & 3.0/3.0 & 500  \\
      &         &   & 4 & 0.1009(10) & 0.1062(9) & (20, 0.5) & 0.9 - 8.0 & 3.3/5.0 & 500  \\  \hline\hline
\end{tabular*}
}
\end{table*}

\subsection{The monopole clusters}

\begin{figure}[htbp]
\includegraphics[width=90mm]{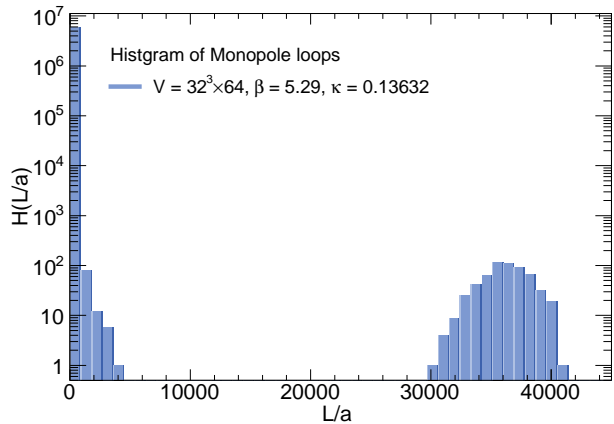}
\caption{The histogram of the length of the monopole loops $L/a$. The normal configurations with two flavors of dynamical Wilson fermions, $V/r_{0}^{4} = 1.565(13)\times10^{3}, \ r_{0}/a = 6.05(5)$, are used.}
\label{fig:mon_clus2}
\end{figure}
\begin{figure}[htbp]
\includegraphics[width=90mm]{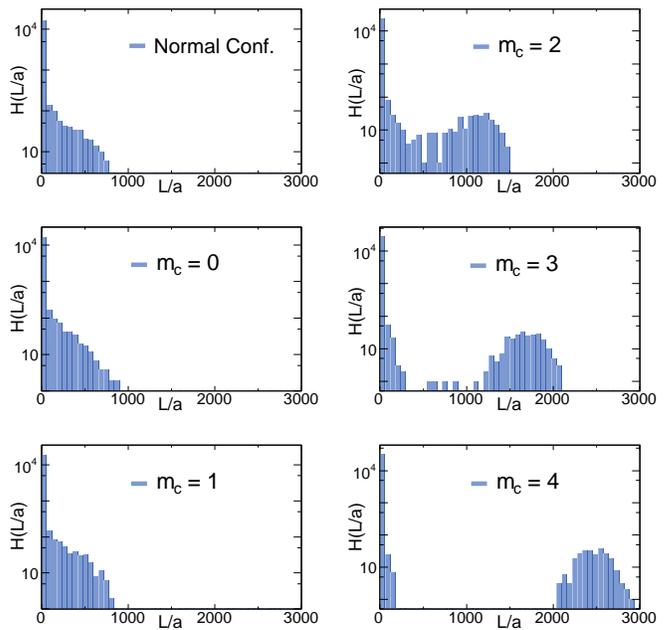}
\caption{The histogram of the length of the monopole loops $L/a$. Monopoles with charges $m_c $ ranging from $0$ to $4$ are added to the normal configurations. The lattice is $V = 14^{4}$, $\beta = 6.00$}
\label{fig:Mon_loops_hist1}
\end{figure}
 The monopoles are known to form two clusters~\cite{Kanazawa2, Kanazawa3, Hart1} in MA gauge. The small (ultraviolet) clusters are made of the short monopole loops, which are short-range fluctuations. The large (infrared) clusters, which percolate through the lattice and wrap around the boundaries, are made of the longest monopole loop in each color direction. The way how to compute numerically the monopole world line in four dimension is explained in~\cite{Bode1}. If the physical lattice volume is large enough, the small clusters and the large clusters are separated as in Figure \ref{fig:mon_clus2}. In quenched SU(2) study~\cite{Kanazawa3, Kanazawa4}, the long monopole loops only exist in confinement phase, and dominate the string tension. The long monopole loops are therefore considered to play the important role of producing color confinement. Going to the Maximally Abelian gauge is essential to divide monopole into clusters.

\begin{figure}[htbp]
\includegraphics[width=90mm]{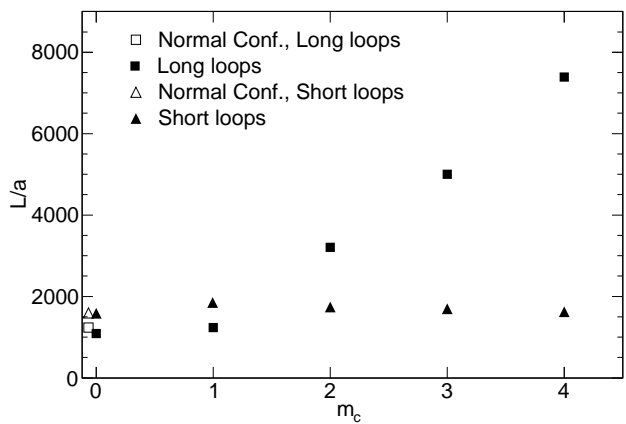}
\caption{The average of the long and short monopole loops divided by the number of configurations. The lattice is $V = 14^{4}$, $\beta = 6.00$. The data points of the normal configurations are slightly shifted from $m_{c} = 0$ for graphical reasons. }
\label{fig:Mon_loops_long1}
\end{figure}

 We create a histogram of the length of the monopole loops when one pair of monopoles with charges from zero to four are added as shown in Figure \ref{fig:Mon_loops_hist1}. To clarify which cluster increases with the monopole charges, we deduct the sum of the longest loops from the sum of all loops. We define the remainder of the subtraction as the sum of short loops. The averages of the sums of the long loops and short loops divided by the total number of configurations are determined. The results are shown in Figure \ref{fig:Mon_loops_long1}. We conclude that the monopoles added are long monopoles wrapping the lattice, as expected.
The length of the long loops is proportional to the charge: a charge $m_{c}$ is equivalent to $m_{c}$ monopole pair of charge 1.

\section{The relations between Zero modes, instantons, and monopoles}

\subsection{Simulation details}

\begin{figure}[htbp]
\includegraphics[width=90mm]{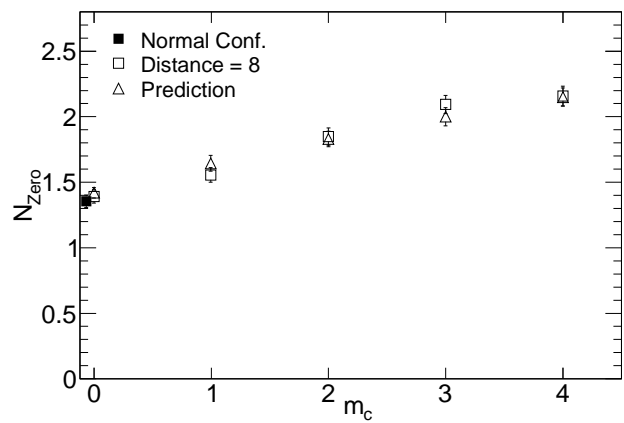}
\caption{The number of zero modes $N_{Zero}$ versus monopole charges $m_{c}$. The data point computed from the normal configurations is slightly shifted from $m_{c} = 0$.}
\label{fig:Add_zero1}
\end{figure}
\begin{figure}[htbp]
\includegraphics[width=90mm]{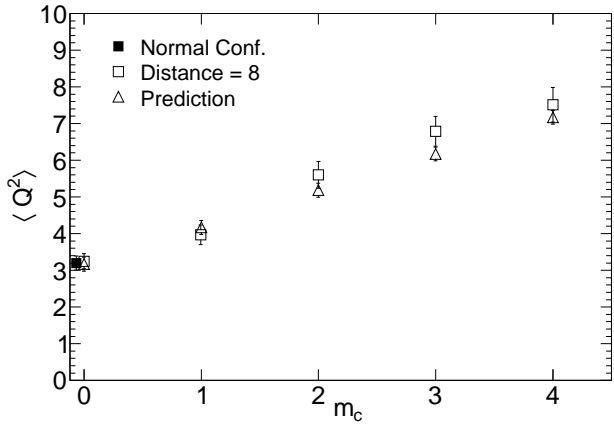}
\caption{The average square of topological charges $\langle Q^{2} \rangle$ versus monopole charges $m_{c}$. The data point computed from the normal configurations is slightly shifted from $m_{c} = 0$.}\label{fig:Add_q21}
\end{figure}

 We generate configurations with one monopole-anti-monopole pair added with different magnetic charges $m_{c}$. The distances between the monopole and anti-monopole are fixed at 6, and 8: slightly changing the distance between them, allows to check finite lattice volume effects. The simulation parameters and the data are listed in Table \ref{tb:Add_zero_1}. The Overlap Dirac operator is constructed from the gauge links of the configurations. The eigenvalue problems are solved by the Arnoldi method, and $\mathcal{O}$(60) pairs of the low-lying eigenvalues and eigenvectors of the Overlap Dirac operator are saved. We then count the number of zero modes, and calculate the average square of topological charges, i.e. the number of instantons. Finally, we compare the simulation results with the predictions based on the hypothesis that the added monopoles do not perturb the vacuum and specifically the distribution of instantons, but can only change the total topological charge.

 Note: We do not do smearing, cooling, or MA gauge fixing in simulations. The number of zero modes, Distance 6 $m_{c}  = 0$, coincide within errors with Distance 8. We take this as a check of volume independence.
\begin{table}
\caption{Charges of the added monopole, together with number of zero modes $N_{Zero}$, and the average square of topological charges $\langle Q^{2} \rangle$. The lattice is $\beta = 6.00$, $V = 14^{4}$.}\label{tb:Add_zero_1}
{\renewcommand{\arraystretch}{1.2}
\begin{tabular*}{\columnwidth}{c @{\extracolsep{\fill}}cccccc} \hline\hline
Distance & $m_{c}$ & $N_{Zero}$ & $\langle Q^{2} \rangle$ & $N_{conf.}$ \\ \hline 
6 & 1  &  1.50(6)  & 3.6(3) & 400 \\   
  & 2  &  1.78(7)  & 5.4(4) & 401 \\ 
  & 3  &  2.03(8)  & 6.7(5) & 423 \\ 
  & 4  &  2.10(8)  & 7.2(5) & 440 \\   
8 & 0  &  1.39(5)  & 3.2(2) & 501 \\ 
  & 1  &  1.55(6)  & 4.0(3) & 502 \\ 
  & 2  &  1.85(7)  & 5.6(4) & 510 \\ 
  & 3  &  2.09(7)  & 6.8(4) & 500 \\ 
  & 4  &  2.16(8)  & 7.5(5) & 505 \\ \hline\hline   
\end{tabular*}
}
\end{table}

 In Figure \ref{fig:Add_zero1} and Figure \ref{fig:Add_q21} we show the number of zero modes and the average square of the topological charges respectively as functions of the monopole charge $m_{c} $. $m_c = 0$ is the ordinary case with no monopoles added. The results are compared with the predictions developed below.

\subsection{Predictions}

 The creation operator Eq. (\ref{m}) acting on the vacuum produces a pair of static monopoles propagating in time from $-\infty$ to $+\infty$. If the vacuum is a dual superconductor it is a coherent state and an eigenstate of the creation operator of the pair, and is therefore left unchanged. In any case, as a dual superconductor the vacuum shields the monopoles, so that they are one-dimensional structures and do not influence the external space time: $O(V^{-\frac{3}{4}})$. For these reasons we expect that the distribution of zero modes Eq. (\ref{n_{+}}), \ (\ref{n_{-}}) is unaffected by the additional monopoles especially at large space-time volumes $V$, except for the addition of some instantons. 

If each monopole-antimonopole pair of charge $m_{c} =1$ produces $q$ instantons or anti-instantons with equal probability ($CP$ invariance), and the original distribution is not modified, the value of $\langle Q^2 \rangle$ becomes
\begin{equation}
\langle (Q + \Delta Q)^2 \rangle = \langle Q^2 \rangle + 2 \langle Q \rangle \langle \Delta Q \rangle _{I} + \langle (\Delta Q)^2 \rangle _{I} \label{DQS}
\end{equation}
Here $\langle \cdots \rangle$ means average on the configuration, $\langle \cdots \rangle_{I}$ the average on the charge distribution of the additional $M \equiv m_{c} \times q$ instantons. The middle term in the right-hand side of Eq. (\ref{DQS}) vanishes since $\langle Q \rangle  =0$. To define $\langle \cdots \rangle_{I}$ we assume that each chiral charge can be $\pm 1$ with equal probability. The probability to have $k$ positive and $M - k$  negative instantons is 
\begin{equation}
P(k, M) = \frac{1}{2^M} \binom{M}{k}, \  (k = 0 ... M).
\end{equation}
Corresponding to this partition $\Delta Q = 2k - M$ so that
\begin{equation}
\langle (\Delta Q)^2 \rangle _{I} = \sum _{0}^ {M} P(k, M) (2k - M)^2 = M.
\end{equation}
The sum is easily computed as shown in Appendix \ref{App1}. Recalling that $M = m_{c} q$, the prediction follows that, in presence of $m_{c}$ pairs of monopoles 
\begin{equation}
\langle (Q + \Delta Q)^2 \rangle  =  \langle Q ^2 \rangle  + m_{c} q \label{deltaq2}
\end{equation}
Comparing with the data [Figure \ref{fig:Add_q21}] it follows that $q = 1$. Fit of a linear function $\langle Q^{2} \rangle = A \cdot m_{c} + B$ to the results of the average square of the topological charges computed from simulations gives a slope $A$ compatible with 1, and intercept $B$ compatible with $\langle Q^{2} \rangle = 3.17 (19)$ (the value from normal configurations) [Table \ref{tb:f_res1}]. 

 The observed behaviour supports the assumption that monopoles do not alter the vacuum. Each pair of monopoles adds one instanton or one anti-instanton with equal probability. Having fixed once and for all that $q = 1$ we can explicitly compute the topological charge distribution, under our basic assumption, and compare with data. Since our monopole configuration is $CP$ invariant, the distribution has to be even in Q. In the simple case $m_{c} = 1$ the distribution is the sum with equal weight $\frac{1}{2}$ of two "unperturbed" distributions $P_{0}(Q)$ centred at $Q = 1$ and $Q = -1$, $P(Q) = \frac{1}{2}[P_{0} (Q+1) + P_{0} (Q-1)]$. $P_{0}(Q)$ is well approximated by Eq. (\ref{eq:pb_func1}). In the case of general $m_{c}$
\begin{align}
 & P(Q, m_{c}) = \nonumber \\ 
 & = \frac{1}{2^{m_{c}+1}}\sum _{k = 0}^{m_{c}} \binom{m_{c}}{k} [P_0 (Q+m_{c}-2k) + P_0 (Q-m_{c}+2k)]\label{eq:prob_func_1}
\end{align}

 The distribution is $CP$ invariant if $P_{0}$ is such.

 By use of Eq. (\ref{eq:prob_func_1}) we can predict $\langle |Q| \rangle$ numerically, using the topological charges computed from the normal configurations as  input value, keeping in mind that pairs of zero modes with opposite chirality escape detection. The computation is done in detail in Appendix section \ref{sec:comp_predi_1} and compared to the data in Figure \ref{fig:Add_zero1}. 

\begin{figure}[htbp]
 \begin{center}
 \includegraphics[width=90mm]{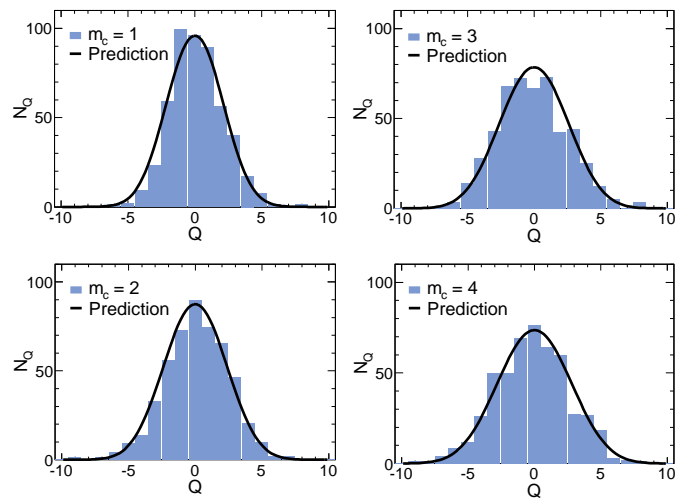}
 \caption{The comparisons of the distributions of the topological charges computed by simulations (Distance = 8) and the prediction. The monopole charges $m_{c}$ are from 1 to 4.}
 \label{fig:top_ch_q1_pred}
 \end{center}
\end{figure}

 The histograms of the topological charges $Q$ for each monopole charge $m_{c}$ are compared to the prediction obtained by multiplying the distribution Eq. (\ref{eq:prob_func_1}) $\times$ the number of configurations in Figure \ref{fig:top_ch_q1_pred}. The input value $N_{I} = 3.17$ is used. The topological charge distributions computed by simulations are represented by the histogram, and the prediction by the continuous line with symbols. The $\chi^{2}/ndf$ indicates agreement with the prediction by Eq. (\ref{eq:prob_func_1}).

 In the general case 
\begin{equation}
\langle Q^2 \rangle = \langle \delta^2  \rangle + m_{c}
\end{equation}
$\langle \delta^2 \rangle$ is nothing but $\langle Q^2 \rangle$ at $m_{c}=0$. Details of the computations are given in Appendix section \ref{sec:comp_predi_1}.

\begin{table}
\caption{The results by fitting the function Eq. (\ref{eq:prob_func_1}).}\label{tb:dis_add_normal_1}
{\renewcommand{\arraystretch}{1.2}
\begin{tabular*}{\columnwidth}{l @{\extracolsep{\fill}}cccc} \hline\hline
             & $m_{c}$ & $N_{I}$ & $\chi^{2}/ndf$ & $N_{conf.}$  \\ \hline
Distance 6   & 1      & 2.5(3) & 8.5/11.0  & 400 \\
             & 2      & 3.3(4) & 9.9/14.0 & 401 \\
             & 3      & 3.3(4) & 15.5/15.0 & 423 \\
             & 4      & 3.0(5) & 8.5/16.0  & 440 \\ 
Distance 8   & 1      & 2.7(2) & 13.6/13.0 & 502 \\ 
             & 2      & 3.4(4) & 8.2/15.0  & 510 \\
             & 3      & 3.6(4) & 14.1/14.0 & 500 \\
             & 4      & 3.1(5) & 16.2/17.0 & 505 \\ \hline\hline
\end{tabular*}
}
\end{table}

\begin{table}
\caption{The final results.}\label{tb:f_res1}
{\renewcommand{\arraystretch}{1.2}
\begin{tabular*}{\columnwidth}{l @{\extracolsep{\fill}}ccccc} \hline\hline
              & $A$ & $B$ & FR ($m_{c}$) & $\chi^{2}/ndf$ \\ \hline 
Prediction    &  1.00(6)  &  3.2(2)    & 1 - 5 & 0.0/3.0 \\
Distance 6    &  1.05(11) &  3.06(18)  & 0 - 4 & 5.7/3.0 \\
Distance 8    &  1.13(10) &  3.15(18)  & 0 - 4 & 2.2/3.0 \\ \hline\hline
\end{tabular*}
}
\end{table}

\section{Summary and conclusions}

 The aim of this paper is to shed light on the relation between instantons and monopoles. 

 Our system is an $SU(3)$ pure gauge theory on the lattice. Our strategy is to add pairs of static monopoles of opposite magnetic charge to the system and detect the consequent variation in the number of instantons.
 
 To add monopoles we use the creation operator developed in Ref.~\cite{DiGiacomo1, DiGiacomo2, DiGiacomo3}. If the vacuum is a dual superconductor we expect that adding the monopoles does not modify the ground state, except in the number of monopoles. We check by the technique of Ref.~\cite{Kanazawa1, Kanazawa2} that we are really adding monopoles: we find that we are indeed adding "long" monopoles, wrapping around the lattice in the (periodic) time direction.

 To detect instantons we count the zero modes of a fermion overlap operator. In doing that we realise that this tool is only able to detect the net topological charge: the zero modes in each configuration are either all of positive chirality, or all of negative chirality, but never appear with opposite sign. The number of observed zero modes is proportional to the square root of the space-time volume $\sqrt{V_{st}}$, and $\langle Q^2 \rangle$ to $V_{st}$ as expected. The only way to reconcile this fact with the obvious expectation that the number of zero modes be proportional to $V_{st}$, the vacuum being translation invariant and the correlation length finite, is to assume that, for some reason, pairs of instantons of opposite chirality escape detection. Maybe this phenomenon disappears in larger lattices. This should be anyhow quantitatively understood, also to have on safe ground the determination of the topological susceptibility. Based on this model we are able to determine the number of instantons and its variation due to the addition of the monopoles.

 We find that an instanton or an anti-instanton are added to the system with equal probability by each pair of monopole-anti-monopole of magnetic charge 1. The variation of the chiral charge is equal to the magnetic charge of the monopole of the pair. This is the our main result.

 We keep running simulations in order to further clarify that Chiral symmetry breaking is induced by the monopole-anti-monopole pair Ref.~\cite{CON}.

\section{Acknowledgments}

 Special thanks are due to F. Pucci, who collaborated to the early stages of this work. M. H. would like to thank sincerely E. -M. Ilgenfritz, Y. Nakamura,  G. Schierholz, and T. Sekido for useful discussions. This study is supported by the Research Executive Agency (REA) of the European Union under Grant Agreement No. PITN-GA-2009-238353 (ITN STRONGnet). M. H. received partial supports from I.N.F.N. at the University of Parma and the University of Pisa. We appreciate the computer resources and technical supports kindly provided by the Research Center for Nuclear Physics and the Cybermedia Center at the University of Osaka, and the Yukawa Institute for Theoretical Physics the University of Kyoto. M. H. would like to thank the STRONGnet Summer School 2011 in Bielefeld for giving him the wonderful opportunity of discussing with F. Pucci about this study. And also, M. H. appreciates the welcoming cordialities from the University of Bielefeld, the University of Kanazawa, the University of Parma, and the University of Pisa.

\appendix

\section{Proof of Eq. (\ref{deltaq2})} \label{App1}

 Consider the function $ f(x)$ defined as
\begin{equation}
f(x)=\sum_{k} P(k, M) x^{k}(\frac{1}{x})^{M-k} = \frac{1}{2^M} (x+\frac{1}{x} )^M
\end{equation}
Compute $ g(1)$ with $g(x) \equiv  x \frac{d}{dx} ( x \frac{df(x)}{dx})$, on both sides of the last equality. The result is Eq. (\ref{deltaq2}).

\section{The computation of $\langle |Q| \rangle$}\label{sec:comp_predi_1}

 From the distribution of topological charge Eq. (\ref{eq:prob_func_1}) we can compute $\langle |Q| \rangle$ as a function of $m_{c}$. $|Q|$ is the observed number of zero modes, being our detection blind to pairs of instantons of opposite chirality. We already know from Eq. (\ref{deltaq2}) that $\langle Q^2 \rangle = \langle \delta^2 \rangle + m_{c}$ where the average $\langle \delta^2 \rangle$ is weighted by $P_{0}(\delta)$, the "unperturbed" distribution.  $\langle \delta^2 \rangle$ is the value at $m_{c}=0$.

\noindent For $m_{c}=1$ we have 
\begin{equation}
N_{Zero}  = \frac{1}{2} \bigl\{ \langle| \delta + 1 | \rangle + \langle| \delta - 1 | \rangle \bigr\}. %= 1.64(6).
\end{equation}
Now,  
\begin{align}
&\langle| \delta + 1 | \rangle \nonumber = \\
& = -\frac{1}{\sqrt{2\pi N}}  \left\{ \int_{-\infty}^{-1}(\delta + 1)e^{-\frac{\delta^{2}}{2N}} d\delta - \int_{-1}^{\infty}(\delta + 1)e^{-\frac{\delta^{2}}{2N}} d\delta \right\} \nonumber \\
& = \sqrt{\frac{2N}{\pi}}e^{-\frac{1}{2N}} - \frac{1}{\sqrt{2\pi N}} \left\{ \int_{-\infty}^{-1}e^{-\frac{\delta^{2}}{2N}} d\delta - \int_{-1}^{\infty}e^{-\frac{\delta^{2}}{2N}} d\delta \right\}\label{eq:nzero1}
\end{align}
\begin{align}
& \langle| \delta - 1 | \rangle  \nonumber = \\ 
& = -\frac{1}{\sqrt{2\pi N}}\left\{ \int_{-\infty}^{1}(\delta - 1)e^{-\frac{\delta^{2}}{2N}} d\delta - \int_{1}^{\infty}(\delta - 1)e^{-\frac{\delta^{2}}{2N}} d\delta \right\} \nonumber \\
& = \sqrt{\frac{2N}{\pi}}e^{-\frac{1}{2N}} + \frac{1}{\sqrt{2\pi N}}\left\{ \int_{-\infty}^{1}e^{-\frac{\delta^{2}}{2N}} d\delta - \int_{1}^{\infty}e^{-\frac{\delta^{2}}{2N}} d\delta \right\}\label{eq:nzero2} 
\end{align}
The number of zero modes $N_{Zero}$ that we observe in our simulation is 
\begin{flalign}
N_{Zero} & = \frac{1}{2} \bigl\{ \langle| \delta + 1 | \rangle + \langle| \delta - 1 | \rangle \bigr\} \nonumber \\
& =  \sqrt{\frac{2N}{\pi}}e^{-\frac{1}{2N}} + \frac{1}{\sqrt{2\pi N}}\int_{-1}^{1}e^{-\frac{\delta^{2}}{2N}} d\delta 
\end{flalign}
The integrals are computed numerically, by substituting $ N =  \langle \delta^{2} \rangle  = 3.17(19)$ of $N_{conf.} = 592$ computed by our lattice simulations. The result is $N_{Zero}  = 1.64(6)$.\\
\noindent For $m_{c} =2$;
\begin{align}
N_{Zero}& = \frac{1}{4}\bigl\{ \langle |\delta + 2| \rangle + \langle |\delta - 2| \rangle \bigr\} + \frac{1}{2}\langle |\delta| \rangle \notag \\
     & = \frac{1}{\sqrt{2\pi N}}\left\{ Ne^{-\frac{2}{N}} + \int_{-2}^{2} e^{-\frac{\delta^{2}}{2N}} d \delta \right\} + \frac{N}{\sqrt{2\pi N}} \nonumber \\
& = 1.82(8)
\end{align}
\noindent For $m_{c}=3$;
\begin{align}
N_{Zero}& = \frac{1}{8}\bigl\{ \langle |\delta + 3| \rangle + \langle |\delta - 3| \rangle\bigr\} + \frac{3}{8}\bigl\{\langle |\delta + 1| \rangle + \langle |\delta - 1 | \rangle \bigr\} \nonumber  \\
& = \frac{1}{8}\biggl\{ \frac{4N}{\sqrt{2\pi N}}e^{-\frac{9}{2N}} + \frac{6}{\sqrt{2\pi N}}\int_{-3}^{3}e^{-\frac{\delta^{2}}{2N}} d\delta  \biggr\} \nonumber \\
& + \frac{3}{8}\biggl\{ \frac{4N}{\sqrt{2\pi N}}e^{-\frac{1}{2N}} + \frac{2}{\sqrt{2\pi N}}\int_{-1}^{1}e^{-\frac{\delta^{2}}{2N}} d\delta   \biggr\}\nonumber \\
& = 1.99(9)
\end{align}
\noindent For $m_{c}=4$;
\begin{align}
N_{Zero} & = \frac{1}{16}\bigl\{ \langle |\delta + 4| \rangle + \langle |\delta - 4| \rangle \bigr\}\nonumber \\  
& + \frac{4}{16}\bigl\{\langle |\delta + 2| \rangle + \langle |\delta - 2| \rangle \bigr\} + \frac{6}{16} \langle |\delta| \rangle \nonumber  \\
& = \frac{1}{16}\biggl\{\frac{4N}{\sqrt{2\pi N}}e^{-\frac{8}{N}} + \frac{8}{\sqrt{2\pi N}}\int_{-4}^{4}e^{-\frac{\delta^{2}}{2N}} d\delta \biggr\}\nonumber \\ 
& + \frac{4}{16}\biggl\{\frac{4N}{\sqrt{2\pi N}}e^{-\frac{2}{N}} + \frac{4}{\sqrt{2\pi N}}\int_{-2}^{2}e^{-\frac{\delta^{2}}{2N}} d\delta \biggr\} + \frac{3N}{4\sqrt{2\pi N}}  \nonumber \\
& = 2.15(9) 
\end{align}

\end{document}